%% file: main.tex
\documentclass[sigconf, nonacm]{acmart}
\usepackage{enumitem}
\usepackage{booktabs}
\usepackage{multirow}
\usepackage{array}

\AtBeginDocument{%
  }

\copyrightyear{2025}
\acmYear{2025}
\setcopyright{cc}
\setcctype{by-nc-sa}
\acmConference[CHI '25]{CHI Conference on Human Factors in Computing Systems}{April 26-May 1, 2025}{Yokohama, Japan}
\acmBooktitle{CHI Conference on Human Factors in Computing Systems (CHI '25), April 26-May 1, 2025, Yokohama, Japan}\acmDOI{10.1145/3706598.3714263}
\acmISBN{979-8-4007-1394-1/25/04}

\begin{document}

\title{SpeakEasy: Enhancing Text-to-Speech Interactions for Expressive Content Creation}

\author{Stephen Brade}
\email{brade@mit.edu}
\affiliation{%
  \department{Electrical Engineering \& Computer
Science Department}
\institution{Massachusetts Institute of Technology}
  \city{Cambridge}
  \state{Massachusetts}
  \country{USA}
}

\author{Sam Anderson}
\email{sanderson@adobe.com}
\affiliation{%
  \institution{Adobe Research}
  \city{New York}
  \state{New York}
  \country{USA}
}

\author{Rithesh Kumar}
\email{ritheshk@adobe.com}
\affiliation{%
  \institution{Adobe Research}
  \city{Toronto}
  \state{Ontario}
  \country{Canada}
}

\author{Zeyu Jin}
\email{zejin@adobe.com}
\affiliation{%
  \institution{Adobe Research}
  \city{San Francisco}
  \state{California}
  \country{USA}
}

\author{Anh Truong}
\email{truong@adobe.com}
\affiliation{%
  \institution{Adobe Research}
  \city{New York}
  \state{New York}
  \country{USA}
}
% \author{Stephen Brade\textsuperscript{*}, Sam Anderson\textsuperscript{\textdagger}, Rithesh Kumar\textsuperscript{\textdagger}, Zeyu Jin\textsuperscript{\textdagger}, Anh Truong\textsuperscript{\textdagger}}

% \affiliation{
%     \begin{tabular}{@{}p{6cm}<{\centering} p{6cm}<{\centering}@{}} 
%         \textsuperscript{*}Massachusetts Institute of Technology & \textsuperscript{\textdagger}Adobe Research \\
%         brade@mit.edu & \makebox[0pt]{\{sanderson, ritheshk, zejin, truong\}@adobe.com} \\
%         \country{USA} & \country{USA}
%     \end{tabular}
% }

\renewcommand{\shortauthors}{Brade et al.}

\begin{abstract}
Novice content creators often invest significant time recording expressive speech for social media videos. While recent advancements in text-to-speech (TTS) technology can generate highly realistic speech in various languages and accents, many struggle with unintuitive or overly granular TTS interfaces. We propose simplifying TTS generation by allowing users to specify high-level context alongside their script. Our Wizard-of-Oz system, SpeakEasy, leverages user-provided context to inform and influence TTS output, enabling iterative refinement with high-level feedback.  This approach was informed by two 8-subject formative studies: one examining content creators' experiences with TTS, and the other drawing on effective strategies from voice actors. Our evaluation shows that participants using SpeakEasy were more successful in generating performances matching their personal standards, without requiring significantly more effort than leading industry interfaces.

\end{abstract}



\begin{teaserfigure}
  \centering
  \includegraphics[width=0.9\textwidth]{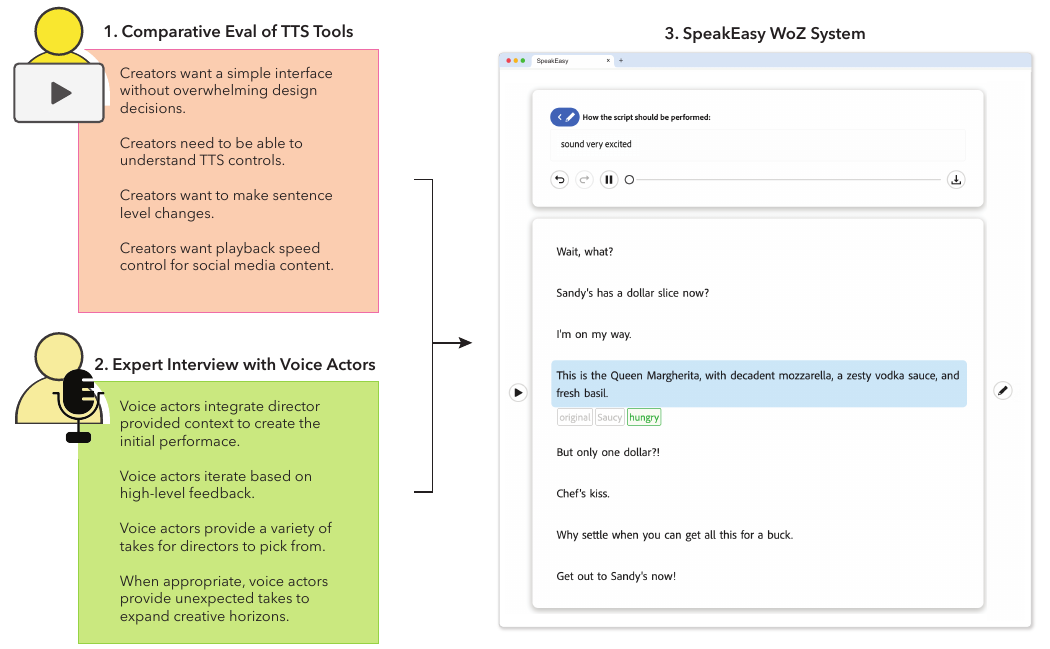}
  \caption{We take a two pronged approach towards designing Text-To-Speech (TTS) interactions for expressive content creation. First, we conduct a comparative evaluation of existing TTS tools with eight independent content creators who record expressive speech for their own videos. From this study, we uncover user needs that are unmet by current tools. Second, we conduct expert interviews with eight professional voice actors to understand effective strategies of communicating intention for expressive speech performances. We synthesize the findings from both studies into a set of design guidelines. We validate these guidelines by reifying them into SpeechEasy, a Wizard-of-Oz system for crafting expressive TTS performances.}
  \Description{A diagram showing the findings from two formative studies with an arrow joining them and pointing to an TTS editing interface to signify that these studies informed the interface. The first study is a comparative evaluation of TTS tools and its findings are: (1) creators want a simple interface without overwhelming design decisions; (2) creators need to be able to understand TTS controls; (3) creators want to make sentence level changes; (4) creators want playback speed control for social media content. The second study is expert interviews with voice actors and the findings are: (1) voice actors integrate director provided context to create an initial performance; (2) voice actors iterate based on high-level feedback; (3) voice actors provide a variety of takes for directors to pick from (4) when appropriate, voice actors provide unexpected takes to expand creative horizons.}
  \label{fig:teaser}
\end{teaserfigure}

\maketitle
\input{sections/1_introduction}
\input{sections/2_related_work}
\input{sections/3_formative_studies}
\input{sections/4_design_goals}
\input{sections/5_system}

\input{sections/6_backend_implementation}
\input{sections/7_user_evaluation}
\input{sections/8_results}

\input{sections/9_discussion_and_future_work}
\input{sections/10_conclusion}

\bibliographystyle{ACM-Reference-Format}
\bibliography{references}

%TC:ignore
\appendix
\input{sections/11_appendix}
%TC:endignore 

\end{document}

%% file: sections/1_introduction.tex
\section{Introduction}
Expressive speech is an essential component of social videos. A high-quality voiceover enables content creators to establish a strong connection with their viewers, add depth and nuance to their stories, and effectively convey their message. However, our interviews show that novice creators often struggle to record quality voiceovers, finding the process to be arduous and time-consuming. To ensure a compelling performance, creators need to develop the skill to create a vocal interpretation that resonates with their audience, rather than simply reading off a script. The challenge lies in finding the right tone, pace, and emotional delivery that will captivate viewers. Moreover, the process of recording quality voiceovers is often plagued by trial-and-error, as creators repeatedly record multiple takes of each sentence to account for any mistakes or imperfections. This laborious procedure involves reviewing every take, selecting the best performance for each sentence, and then splicing these performances together into the final voiceover. The result is a time-consuming process that can be detrimental to social media creators who need to publish content quickly.

Recent breakthroughs in text-to-speech (TTS) technology offer a promising solution to alleviate the time-consuming burden on content creators. 
State of the art (SOTA) TTS algorithms are capable of generating remarkably realistic speech in a multitude of voices, and a wide variety of accents and languages~\cite{choice_of_voices, believing_anthropomorphism}. These algorithms underpin voice assistants \cite{conversational_user_interfaces} \footnote{Apple's Siri: \url{https://www.apple.com/ca/siri/}, Amazon's Alexa: \url{https://developer.amazon.com/en-US/alexa}}, support translation software \footnote{Google Translate: \url{https://translate.google.ca/?sl=en&tl=it&op=translate}} and enable new accessible features for the visually impaired \cite{voicesetting, blind}. In addition to alleviating recording challenges, TTS has the potential to democratize voiceover production for individuals with speech impediments, accents, or anonymity concerns.

Despite its promising capabilities, current TTS technology is hindered by several constraints that limit its practical usability for content creators. A key constraint is the lack of control over outputs; past research has underlined a critical need for creating synthetic speech that accurately conveys emphasis, prosody, and broad emotional textures~\cite{unlocking_creator_ai_synergy, speechless, voicesetting}. Studies also have shown that emotional incongruity is a common point of distinction between real and synthetic speech \cite{uncovering_human_traits}, and that TTS is less effective than humans at performing persuasive content \cite{uncanny_valley}. Given these limitations, we hypothesize that current TTS tools lack controls that align with the mental models of users authoring expressive speech. This is particularly evident in the need for nuanced deliveries that are tailored to specific video content and creator intentions.

To investigate our hypothesis around expressivity and TTS tools, we conducted a comparative evaluation of two industry leading TTS interfaces among eight creators who specialize in crafting narrative content for social media platforms. 
One interface offered global hyperparameter-based controls, which allowed users to manipulate variables such as variability, while the other provided local control over quantitative qualities of speech, including speed, pitch, and pause length.
In this study, creators were tasked with using each TTS interface to generate synthetic speech for their own scripts from a past video. Their objective was to produce a satisfactory performance, but not to exactly replicate their own performance. Our findings revealed that the hyperparameter-based controls proved perplexing to users, who found them difficult to interpret and understand. As a result, participants often resorted to trial-and-error approaches, randomly exploring these control parameters in search of a desirable outcome. In contrast, creators found the quantitative speech quality controls more intuitive and easier to grasp. However, they also reported feeling overwhelmed by the sheer number of design decisions required to achieve a satisfactory performance using these controls. Notably, despite their frustration with the manual recording process, six out of eight creators concluded that generating expressive speech with TTS requires significantly more effort and time than manually recording the performance themselves.

To inform the development of more effective TTS control systems for expressive speech generation, we interview eight professional voice actors to understand how directors communicate their intentions, and how voice actors interpret those intentions to deliver expressive speech performances. Our findings reveal that directors usually give voice actors an initial context to accompany the script, and provide high-level feedback that is often vague and emotional to guide performance iterations. Voice actors then either provide a wide variety of suitable performances within these loose boundaries, or explicitly push beyond the boundaries to expand the creative possibilities of the project.

We distill the insights from our comparative study and voice actor interviews into a cohesive set of design principles for crafting expressive TTS interfaces. 
To validate these guidelines, we implemented SpeakEasy, a Wizard-of-Oz TTS interface designed to facilitate creative speech generation. The Wizard-of-Oz approach allows us to quickly disseminate insights on interactions that will be made possible with emergent technologies\cite{GPT4o} not widely available at the time of writing and to inform the development of these technologies. SpeakEasy incorporates user-provided context to influence TTS output, and allows creators to iterate on individual sentence performances with high-level feedback. SpeakEasy also employs a recommendation feature that suggests a diverse range of possible speech performances for the user’s script, enabling users to easily explore and compare different performances. 
We evaluated SpeakEasy in a within-subjects study with twelve video creators who compared our system to the two baseline interfaces used in the formative study. The results showed that users are significantly more successful at generating speech that reaches their personal standards when using SpeakEasy compared to both baselines. 

In summary, our work make the following contributions:
\begin{itemize}
\item A comparative evaluation with eight narrative-media content creators that identifies the common workflows and pain-points of using two industry leading TTS systems to generate expressive speech.
\item An interview study with eight professional voice actors that uncovers how humans communicate and interpret intentions for performing expressive speech.
\item A Wizard-of-Oz system, SpeakEasy, that leverages high level user feedback as input to contextually generate expressive TTS content of a given script.
\item A comparative evaluation which demonstrates the effectiveness of SpeakEasy, and provides insights towards the development of future TTS algorithms and interfaces that prioritize expressiveness.
\end{itemize}

%% file: sections/2_related_work.tex
\section{Related Work}
\subsection{Text-to-Speech Algorithms}
Text-to-speech refers to a class of algorithms that can intake a script, and output an understandable waveform that resembles human speech. Early approaches focused on applying digital signal processing (DSP) to synthesize speech \cite{dutoit1997high}. Rapid advancements came with the application of neural networks to audio synthesis like WaveNet \cite{van2016wavenet}, which presented a deep architecture for raw waveform generation, and later TacoTron \cite{Wang2017TacotronTE}, which achieved end-to-end single-speaker speech synthesis by leveraging a sequence-to-sequence approach. More recent zero-shot approaches are capable of generating speech in a particular voice without explicitly training on that voice \cite{Betker_TorToiSe_text-to-speech_2022, borsos2023soundstorm, valle, valle2}. Other approaches have focused on adding emotional conditioning to TTS models in order to generate speech of a particular emotional quality \cite{overview_of_affective}. However, past scholarship has highlighted limitations of these approaches, particularly their inability to generate speech that is persuasive, or suitable for attention grabbing content like sports broadcasting \cite{unlocking_creator_ai_synergy, speechless, voicesetting, human_all_too_human, uncovering_human_traits, uncanny_valley}. Recent work has attempted to allay these limitations by enabling prompt-based generation of speech for a particular gender, speed, pitch, volume or vocal timbre \cite{guo2022promptttscontrollabletexttospeechtext, leng2023prompttts2describinggenerating, liu2023promptstylecontrollablestyletransfer}. However, these prompting techniques are still constrained to descriptions of low-level speech attributes (e.g. gender, pitch, etc.) that users may struggle to articulate with respect to their creative intentions. More recently, GPT-4o Voice~\cite{GPT4o} has demonstrated remarkable capabilities for expressive speech generation given a high-level user prompt\footnote{\url{https://www.youtube.com/watch?v=D9byh4MAsUQ}}. At the time of writing, highly expressive technologies like GPT-4o Voice were broadly unavailable. Therefore, in SpeakEasy we leverage a Wizard-of-Oz approach to simulate an expressive TTS system to provide timely design guidelines for applying such TTS models to support creative speech generation.

\subsection{Interfaces for Text-to-Speech Algorithms}
TTS algorithms have been wrapped in a variety of interfaces to serve a diversity of needs. It has been used in conversational user interfaces (CUI) to enable voice assistants in scholarly work \cite{cui, conversational_user_interfaces} and in industry (e.g. Amazon's Alexa, Apple's Siri, etc.). TTS has also been repackaged to enhance video conference call experiences, enabling users to speak when they are in an overly loud, or strictly silent environment \cite{i_cant_talk_now, speechless}. It has been used as a foundation for video prototyping \cite{video_prototyping} and explored for enhancing accessibility for users with vision-impairment \cite{blind}, reading disabilities \cite{dyslexic}, or speech disabilities \cite{voicesetting}. Voicesetting \cite{voicesetting} in particular explores the rapid addition of expressivity to synthesized speech via their Expressive Keyboard, and a slower but more precise addition of expressivity through the Voicesetting editor. This body of research, as well as Voicesetting, contributes bottom-up approaches that build on existing technology. We compliment these works by exploring an idealized Wizard-of-Oz system to create goal posts for future advancements in interface design for TTS.

\subsection{Speech Authoring Interfaces}
A breadth of work in human-computer interaction (HCI) has focused on developing systems that facilitate speech authoring. Some interfaces focus on providing users with feedback on where and how they should modify their speaking tone to record more engaging speech. For example, VoiceCoach \cite{wang2020voicecoach} develops an interface which provides users with semantically relevant examples from TED Talks \cite{TedTalks} that can inspire a new speech pattern. Narration Coach \cite{capture_time} provides textual feedback on how well a user is following professional recommendations for low-level features like speed, pitch, and volume. Other designs have focused on exploring techniques that enhance self-perception of speech \cite{voice_manipulation}, or improve the intonation of non-native English speakers \cite{verbose}. Related work has also focused on improving audio quality by identifying, and automatically dubbing sections of speech with low quality \cite{autodub}, or providing a human-machine feedback loop that improves audio quality by eliminating background noise, and optimizing reverb \cite{autodub}. Whereas these works focus on providing pathways for improving recorded speech, we aim to provide pathways that help users improve synthesized speech. Although these two research areas share a common modality, these systems will need to be optimized in different ways to enhance user experiences. For speech authoring interfaces, the interface must be designed to give actionable feedback to the user. For TTS systems, the interface must be designed to incorporate feedback that comes to the user most easily. It is unclear from the outset whether feedback that users find actionable is also intuitive for them to give to a TTS interface. We address this issue by learning typical feedback styles that directors give voice actors, and incorporating these feedback styles into our system.

%% file: sections/3_formative_studies.tex
\section{Comparative Evaluation of Existing TTS Interfaces}
\label{sec:formative-compare}
\begin{figure*}[t]
  \centering
  \includegraphics[width=\linewidth]{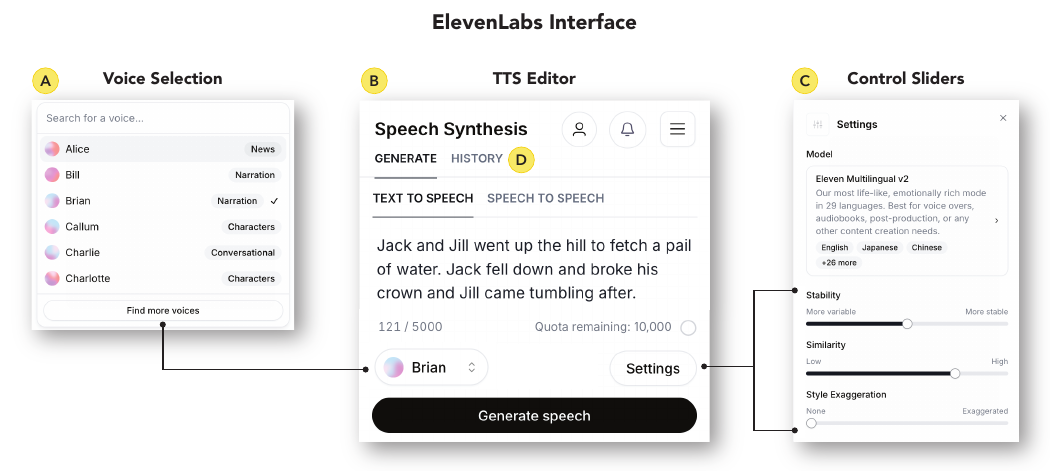}
    \vspace{-1cm}
  \caption{The ElevenLabs interface contains \textbf{(A)} a drop-down menu for voice selection; \textbf{(B)} a text box editor for inputting the script and generating speech; \textbf{(C)} a control panel with sliders for stability, similarity, and style exaggeration, which control model hyperparameters and model stochasticity\protect\footnote{\url{https://elevenlabs.io/docs/speech-synthesis/voice-settings}}; \textbf{(D)} a history tab for revisiting past generations.}
  \label{fig:elevenlabs}
  \Description{A diagram showing the web browser interface for the text-to-speech service ElevenLabs. There are three sections from left to right which are labelled A, B, and C. The central browser (labelled as B. TTS Editor) contains a text box in which a nursery rhyme is entered as example script. It contains a two tabs at the top, labelled "Generate", and "History". History is described in the figure caption and further labelled as "D". At the bottom of the central panel, there are three action items. 1) a button which says "Generate Speech", 2) a drop-down menu for voices currently set to "Brian", and 3) a button labelled as "Settings". Attached to the drop-down menu with a line is a screenshot of the opened drop-down menu (labelled as A. Voice Selector) for voice selection which contains a list of names (e.g. Alice, Bill, Brian, etc.) and labels (e.g. news, narration, characters, etc.). Attached to the settings button with a black line is the opened settings menu (labelled as C. control sliders). The black line diverges to highlight three control sliders for controlling speech generation called "stability", "similarity", and "style exaggeration".}
\end{figure*}

\begin{figure*}[h]
  \centering
  \includegraphics[width=\linewidth]{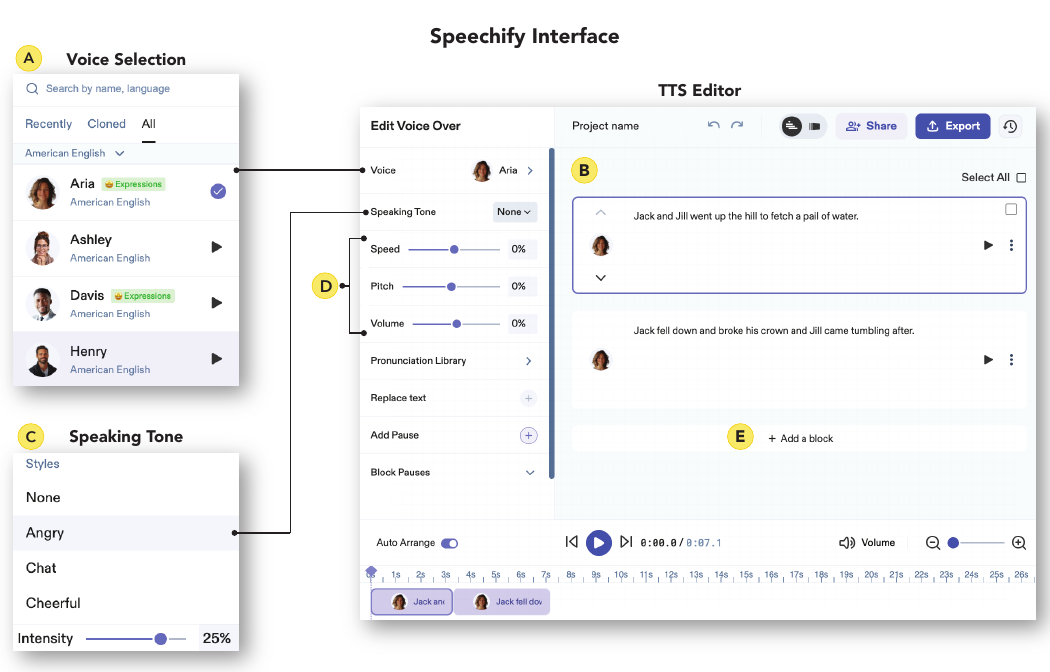}
  \vspace{-1cm}
  \caption{The Speechify interface contains \textbf{(A)} a drop down menu for voice selection with regional accent labels and avatars; \textbf{(B)} a text box editor for inputting the script and generating speech; \textbf{(C)} a tone control drop-down menu (for some voices) which contains a list of emotions, and an intensity slider which changes the tone with respect to a given emotion and intensity; \textbf{(D)} granular sliders for speed, pitch, and volume control as well as additional controls to change pronunciation and manipulate pauses; \textbf{(E)} the text block feature which allows users to parse the text, and apply different controls to different parts of the script. }
  \label{fig:speechify}
  \Description{This figure contains three screenshots of Speechify features. The largest screenshot on the right is the default view of the speechify editor. The screenshot at the top left is the voice selection drop down menu. The screenshot at the bottom left is the speaking tone dropdown menu. Both screenshots on the left are attached to the main editor on the right to indicate where a user must click to access these panels. 5 features are labelled across these three screenshots and labelled as A through E. Feature A labels the voice selection drop down screenshot and contains a list of names and their accents (e.g. Aria, American English). Feature B labels the text box where the script can be entered on the rightmost screenshot of the main interface. Feature C labels the screenashot of the speaking tone with menu items labelled with emotions (e.g. angry, chat, cheerful) and slider to control the intensity of that emotion at the bottom. Feature D labels three sliders for speed, pitch, and emotion. Feature E labels a button for adding a text block.
}
\end{figure*}

To understand the workflow and challenges of creating expressive speech with existing TTS tools,we conducted a comparative evaluation of two widely \cite{ElevenLabsUsers, SpeechifyUsers} used TTS interfaces: ElevenLabs \cite{ElevenLabs} and Speechify \cite{Speechify}. We chose these interfaces for their disparate approaches to user control; ElevenLabs (Figure~\ref{fig:elevenlabs}) provides global, hyperparameter-based control while Speechify (Figure~\ref{fig:speechify}) provides local control over quantitative qualities of speech (e.g., speed, pitch, pause length).

\label{formative_study_methodology}
\subsection{Participants and Procedure}

We recruited eight content creators with at least one year of experience recording scripted speech content for their own social media videos (Appendix \ref{appendix-formative} Table \ref{table:content_creators}). Participants were recruited through User Interviews\cite{UserInterviews} and word-of-mouth. With the exception of C4, most of these creators are not video professionals, but rather use video as a means towards an end-goal (e.g., advertising their business) and feel pressure to produce content quickly and continuously. They record and edit their own content in small teams of one to three people. C4 produces documentaries for a company, but separately also makes their own personal content. Six participants had minimal experience using TTS tools prior to the study.

Each study was conducted over video call and lasted about 60 minutes. The study began with a 10 minute opening interview about their manual speech recording workflows. Participants then used ElevenLabs and Speechify to generate speech for their respective scripts. In order to increase participant investment into the output quality, we asked participants to provide 30-second scripts they previously recorded for their own social media. Their goal was to create a satisfactory performance, but not to exactly replicate their own performance. Participants spent up to 20 minutes in each tool, including a 10 minute run-through of tool features before starting the task. We counterbalanced the order of the tools across participants. We closed the study with a 10 minute, semi-structured interview about the participants' experience using the tools. Participants were compensated \$60 USD.

\subsection{Manually Recording Expressive Speech}
 Participants detailed their current process for manually recording and editing speech and described it as ``very annoying'' and ``takes a lot of time''. Some participants prepare by memorizing the script and contemplating the emotional nuance they want to achieve during their performance. Two participants use costly equipment like specialized microphone or room treatments to ensure high audio quality. Once they start recording, all participants reported needing to re-record multiple takes to achieve a desirable performance. These multiple takes account for mispronunciations, background noise or feedback from other stakeholders. C2, who records several versions of each sentence and then compiles his favorite takes into a final performance, called this process ``one of the biggest pains'' of content creation and that he would ``absolutely use the AI text-to-speech if the quality was there''. 

\subsection{Workflows for Generating Expressive Speech with TTS}

The workflows participants employed while generating expressive speech with TTS consisted of three major steps: (1) voice selection, (2) generating an initial performance and (3) reactive iteration. 

\subsubsection{Voice Selection.} 
Across both tools, all participants began by selecting a voice (Figures ~\ref{fig:elevenlabs}.A, ~\ref{fig:speechify}.A). Six participants wanted to affirm their identities or branding by selecting a voice similar their own in qualities such as accent, cadence, deepness and more. However, five of those participants found this affirmation difficult to achieve with a synthetic voice.

\subsubsection{Generating an Initial Performance.} 
After selecting a voice, participants generated an initial performance of their script. 
Notably, most participants used the default settings and did not proactively adjust any controls before the first generation.
In ElevenLabs, users did not know how to interpret the sliders (Figure ~\ref{fig:elevenlabs}.C) with respect to their intentions and all users left them as is to establish a baseline.
% However, one user did split up the script by adding per sentence line breaks.
Speechify’s controls were more understandable but also required more work to break up the script and change parameters for each chunk. 
Most participants wanted to test whether an initial generation without this added effort would be satisfactory. Only two participants proactively split up the script into several blocks (Figure ~\ref{fig:speechify}.E) and adjusted the tone sliders (Figure ~\ref{fig:speechify}.B) to give some blocks a particular emotional affect.
One participant, C3, felt that their initial ElevenLabs generation was immediately usable for their content and did not move on to the next step for that tool.

\subsubsection{Reactive Iteration.} Other than C3's ElevenLabs result, participants found the initial performances unsatisfactory and spent the majority of their time reactively iterating on the generated speech. After each generation, participants determined what they wanted to change (e.g., speaking speed or tone for certain parts of the script) and adjusted the interface controls to align the generated speech to their creative intentions. In ElevenLabs, participants randomly manipulated the control sliders or added punctuation to the text to change the generation. In Speechify, participants extensively used the text blocking feature and applied different controls to each block. Participants commonly split their script up by individual sentences or by chunks of sentences sharing the same emotional quality. Most users underwent numerous rounds of iteration in both tools because their adjustments did not always produce an output that they liked or expected.

\subsection{Challenges of Generating Expressive Speech with TTS}
While creators were excited about the potential of TTS to speed up their workflow, they encountered a number of challenges in both tools that impeded their experience. Despite expressing frustration at the laborious and time-consuming manual recording process, six of the eight participants reported that using ElevenLabs to generate an expressive performance requires more time and effort compared to manual recording. All eight of the participants perceived that using Speechify requires more time and effort compared to manual recording. Below we report the challenges of generating expressive speech with the current TTS interfaces.

\subsubsection{Controls were difficult to interpret.}
Multiple participants felt confused by ElevenLabs' control sliders. ElevenLabs provides three sliders for "Stability", "Similarity", and "Style Exaggeration" which control model hyperparameters and stochasticity (further descriptions in Appendix \ref{appendix-elevenlabs}). They found it difficult to anticipate how or why the speech would change in a certain way when moving the sliders. Moreover, setting a slider to a specific value did not always yield the same output each time. In contrast, all participants were able to easily interpret the Speechify speed, pitch, and volume sliders. However, half of the participants struggled to interpret the Speaking Tone controls (Figure ~\ref{fig:speechify}.A) since activating those controls did not always produce a noticeable change in tone. 

\subsubsection{Users could not make sentence level modifications.}
When using Speechify, almost all of the participants leveraged the text blocking feature to split up their script by sentence. Conversely, several participants struggled with and explicitly inquired about the lack of sentence level control in ElevenLabs. Participants wanted to make local changes, but editing certain parts of the text or modifying a slider would change the generation for the entire script. To overcome this frustration, C4 even created their own text-blocking strategy by generating a performance per sentence and downloading each audio file independently. 

\subsubsection{Users feel overwhelmed with too many options.}
While participants generally appreciated being able to operate at a sentence level in Speechify, six participants experienced ``decision fatigue'' from the large quantity of granular controls that the tool offered. Iterating in Speechify involved making decisions on multiple parameters such as speed, pitch, volume, pause length, tone or even pronunciations for individual words. This abundance of choice felt especially ``overwhelming'', ``labor intensive'', and ``not worth it'' to participants trying to publish content quickly. In comparison, two participants who were quickly able to produce satisfactory results in ElevenLabs felt that the interface was more enjoyable to use due to its simplicity.

\subsubsection{Playback speed is important for social media.}
Almost all participants used the speed controls provided in Speechify, while multiple criticized the lack of speed control in ElevenLabs. Participants expressed that altering speech speed is particularly crucial for social media content creation. C2 often speeds up his short-form content because ``you're always looking for ways to make it shorter'', while C4 emphasized the importance of ``slow[ing] down on on titles so that people register them and remember them.'' 

\subsubsection{TTS voices sound unnatural.}
Several participants reported that the TTS voices generally sounded ``robotic'' or ``unnatural'', expressing that it was difficult to hear ``the tone or the emotion behind the language'' in both interfaces. Some participants thought that the ElevenLabs voices sounded ``more human'' than the Speechify voices. For two of those participants, the ElevenLabs performance quality was good enough to use in their own content unconditionally. The remaining six participants would not use TTS for their own content, or would only use it to quickly produce content when money is tight.

\subsection{Summary}
Neither interface provided participants with the ability to control TTS in a manner that matches their conception of expressive speech.
This mismatch manifested itself through contrasting experiences with ElevenLabs and Speechify.
Both ElevenLabs and Speechify failed to motivate proactive control to help participants tailor their starting point to their creative vision. As a result, participants spent the bulk of their time reactively iterating. When iterating with ElevenLabs, participants felt frustrated that they could not break up text, modify speed, or not understand the control set. However, at times, ElevenLabs produced very natural speech and some participants were satisfied with their result with little iteration. With Speechify, participants appreciated being able to block text, control speed, and understand most of the controls offered. However, Speechify's controls were overly granular, leading to an overwhelming number of design decisions.

\section{Voice Actor Interviews}

The comparative evaluation study validated our hypothesis that current TTS controls do not match the mental models of users authoring expressive speech. 
To inform alternative approaches to TTS control for expressive speech generation, we conducted interviews with eight professional voice actors. These voice actors provide an expert perspective on how people communicate and interpret intention around nuanced speech performances. 

\subsection{Participants and Procedure}
We recruited eight professional voice actors (Appendix ~\ref{appendix-formative} Table~\ref{table:voice_actors}) to participate in a 45-minute semi-structured interview over video call. We recruited these voice actors through internet forums (i.e., Discord\cite{Discord} and Reddit\cite{Reddit}) and word-of-mouth. All participants had at least two years of professional experience narrating advertisements, explainer videos, audio books and more as their part or full-time job. During the interview, we ask questions about what instructions they are given when preparing for a performance, how they interpret these instructions, and how they incorporate iterative feedback from key stakeholders (i.e. directors). We encouraged them to give concrete examples where possible. Participants were compensated \$60 USD.

\subsection{Instructions for Initial Performances}
In addition to the script, voice actors receive a myriad of materials that help them understand the performance context when preparing for a role. These contextual materials can include character descriptions, celebrities to emulate, example recordings, and/or the video that the speech will accompany. Voice actors integrate this contextual information and employ acting techniques to craft a performance. 
V4 incorporates the script and character descriptions using the Stanislavski Technique \cite{Stanislavski} to ``[find] out what that character wants, what they're prepared to do to get it.'' V8 matches their performance with the video materials to ensure that they ``flow'' and are ``in synchrony''. When instructions are broad (e.g, sound like Jim from the Office), V1 asks the director deeper questions to tease out their intentions (e.g., what just happened to Jim?).

\subsection{Iterating with High Level Feedback}
Voice actors integrate the supplied context to provide an initial performance or set of performances. Directors listen to the performances and give iterative feedback. Most of the voice actors described this feedback as minimal. Directors often ask for another take or give high-level instructions such as ``we want this to sound more professional'', ``could you say it angrier'', or ``can you pace it a little faster''. Some voice actors expressed a preference for specific quantitative feedback (e.g., related to pitch, timbre, etc.) but noted that such feedback is ``harder to give because it's more detailed'' and ``takes more experience to give''. 

\subsection{Providing a Variety of Performances}
Directors often ask voice actors for multiple takes for both the initial performance and subsequent iterations. In some cases, they may not know what they want or may have a hard time articulating it; these options help directors discover what they want. Directors may also ask for variation as a safety to account for project direction changes that they didn't anticipate.
On the voice actor side, providing calculated variations enables them to increase the likelihood of matching the director's creative intention. In V1's experience, the best way to respond to director feedback is to ``to give them options... based on what [they] said, here are three different options that are even tighter into that funnel.''

\subsubsection{Expanding Creative Horizons}
Variations also enable voice actors to contribute creatively by providing performances that the directors would not have otherwise expected. Some voice actors provide ``ABC takes'' where the A, and B takes are more standard interpretations, and the C take is ``over the top''. These C takes could have a different emotional tone or even improvised dialogue to help directors expand their creative horizons. Some directors explicitly ask for these unexpected takes while others are often pleasantly surprised, responding ``we didn't know that was what we wanted, but now that's what we wanted'' (V8). However, voice actors note that this improvisation is only appropriate in certain contexts (e.g., not in corporate contexts where the script has gone through several rounds of approval).

\subsection{Summary}

From these expert interviews, we learned that in addition to the script, voice actors assimilate context from directors to more effectively approximate a satisfactory initial performance. Voice actors then iterate based on high-level and often minimal feedback from directors. They manage this process by providing a variety of performances within the bounds of the feedback. When appropriate, voice actors also help directors uncover more creative possibilities by suggesting variations outside the bounds of the feedback.

%% file: sections/4_design_goals.tex
\section{Design Goals}
From the comparative evaluation, we found that creators appreciated the playback speed and sentence-level controls provided by Speechify, and the minimal and manageable control set provided by ElevenLabs. However, creators struggled to generate an expressive performance in both tools due to Speechify's overwhelming granularity and ElevenLabs' ambiguity. From the voice actor interviews, we learned best practices for communicating and interpreting intentions around expressive speech performances. Specifically, voice actors intake contextual materials from the director and use this context to more effectively approximate a satisfactory initial performance of the script. In contrast, current TTS tools do not help users tailor their starting point and often yield unsatisfactory initial outputs that require great iteration. Voice actors iterate based on high-level, minimally descriptive feedback from directors. They respond by giving a wide variety of takes to help directors narrow down what they want. When appropriate, voice actors contribute creatively by giving improvised and unexpected takes to help directors expand their creative horizons. 

We aggregate the strengths of both the current TTS tools and voice actor workflows to inform the following design goals:
\begin{itemize}
    \item [D1. ] \textbf{Intake user context with the script to intelligently generate an initial performance.} We aim to let users proactively control TTS algorithms by allowing them to include additional context before generating their first performance. This allows users to tailor the initial output to their artistic intention, enabling them achieve a more suitable result with little iteration.
    \item [D2. ] \textbf{Allow for high-level user feedback to iterate on individual sentences.} Upon receiving an initial performance, we aim to let users incorporate high-level feedback (e.g. an adjective, an emotion, etc.) at the sentence-level. High-level feedback will allow for easier iteration that is not overly granular or confusing. Sentence-level control will allow users to modify only the parts that need to be improved.
    
    \item [D3. ] \textbf{Ensure a wide variety of performances and allow for easy comparison of performances.} Ensuring a wide variety of performances will mean that users can get different performances with minimal changes to the control configuration. This variability increases the likelihood that a user will find at least one suitable performance.
    \item [D4. ] \textbf{Expand the user's creative horizons.} We aim to generate performances that expand a user's creative horizon. By providing the option of an unexpected performance, we expose users to performances that they didn't originally consider but may appreciate.
    
    \item [D5. ] \textbf{Enable playback speed control.} We ensure that users can manipulate the speed of playback. In doing so, participants can easily change the length of their speech to fit within, or fill a particular time requirement.
\end{itemize}

%% file: sections/5_system.tex
\section{SpeakEasy System}
To validate our design goals, we operationalize them into a TTS system called SpeakEasy. SpeakEasy is a Wizard-of-Oz system that leverages user-provided context to influence TTS output and enables iteration with high-level feedback. To demonstrate how SpeakEasy works in practice, we walk through a scenario where a creator, Simon, wants to generate the speech track for an Instagram Reel advertising his family's pizza restaurant.

\begin{figure*}[h]
  \centering
  \includegraphics[width=\linewidth]{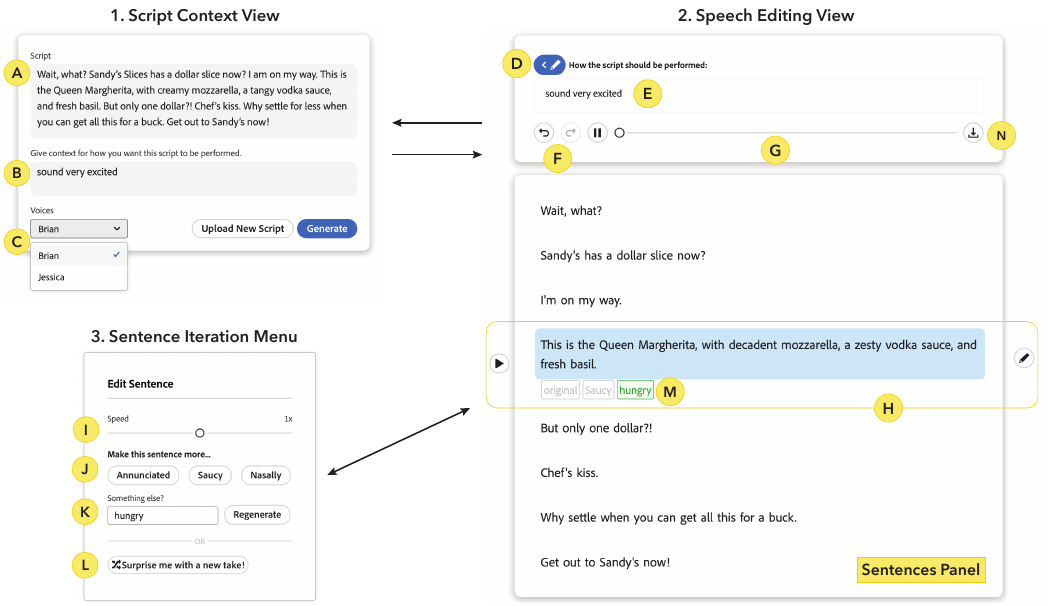}
  \caption{ We present a labelled version of the Wizard of Oz interface, SpeakEasy, with brief descriptions of each feature's functionality. \textbf{1. Script Content View:} \textbf{(A)} \textit{Script Input:}  Users upload a script by loading a text file which is then displayed in this window. \textbf{(B)} \textit{Additional Context Input:}  Users can enter additional context to how they would like the script to be performed. \textbf{(C)} \textit{Voice Selection:} Users can select one of two voices: Brian or Jessica. After the user clicks generate, they are taken to the Script Editing View.  \textbf{Script Editing View:} \textbf{(D)} \textit{Edit context button: }a back button so users can at any time return to the initial context menu to try a new prompt. \textbf{(E)} \textit{Current context display:} Users can see the context they've provided or view the suggested context. \textbf{(F)} \textit{Undo and Redo:} Users can undo any change at any time. \textbf{(G)} \textit{Playback bar:}  Participants can drag this slider to any word in the generated speech. \textbf{(H)} \textit{Active Sentence:} The sentence currently being edited in the Sentence Iteration Menu. \textbf{3. Sentence Iteration Menu:} A menu containing features that help users modify a given sentence. \textbf{(I)} a playback speed slider.\textbf{ (J)} \textit{Adjective Recommendations} adjectives suggested to users based on the current sentence which, when selected, modify the generated speech to reflect this adjective. The first two are fitting, and the third adjective contrasts the first two. \textbf{(K)} \textit{Freeform Text Input:} allows users to modify a sentence with any text that comes to mind. \textbf{(L)} \textit{Surprise Take:} a button which will give the user a completely new performance at random. \textbf{(M)} \textit{Comparison Tabs:} tabs labelled with the word used to edit the sentence that allow users to access past iterations of that sentence, and compare it to the current iteration. \textbf{(N)} a download button to save a performance.}
  \label{fig:interface}
  \Description{This figure contains three numbered screenshots of the SpeakEasy interface connected by arrows to show interface navigation and containing 14 features labelled A through M. The screenshot at the top left is labelled "1. Script Content View" and has three features A-C. Feature A is a textbox where the uploaded script is displayed. The script is a commercial for a pizza restaurant which is advertising their dollar slices. Feature B is the textbox for inputting additional context beyond the script containing "sound very excited". Feature C is a dropdown menu for selecting one of two voices: either brian or jessica. The screenshot on the right is labelled as "2. Speech Editing View" with features labelled D through H and M through N. Feature D is a blue back button to navigate back to the features from "1. Script Content View". Feature E displays the additional context. Feature F displays undo and redo buttons. Feature G is a slider for playback with a pause button to the left. Below these features is a panel containing the script. Feature H is a golden rectangle around one of the sentences which captures sentence level features and has an arrow which connects it to the third screenshot labelled "3. Sentence Iteration Menu". Feature M is the history tabs with three tabs labelled "Original, Saucy, and Hungry" respectively. Feature N is a download button. The screenshot labelled "3. Sentence Iteration Menu" contains three features labelled I through L. Feature I is a slider controlling speed with a label of "1x" to the top right. Feature J are the Adjective Recommendations with three buttons corresponding to adjectives labelled as "Annunciated, Saucy, Nasally". Feature J is a textbox for freeform text input. Feature L is the surprise take feature which is labelled with a shuffle icon and "surpries me with a new take". }
\end{figure*}

\subsection{Script and Initial Context Input} 
Simon starts by uploading a text file of his script into SpeakEasy. The Script Context View (Figure \ref{fig:interface}.1) displays the script (Figure \ref{fig:interface}A) alongside the context text box (Figure \ref{fig:interface}B) where users can specify a high level context for how they want the script to be recited (D1). SpeakEasy supports flexible contextual prompts; users can input high-level instructions (e.g., ``sound very excited'') or character descriptions (e.g., ``you missed lunch and you're famished'') depending on their needs. If users have no context in mind, SpeakEasy will recommend a performance context based on the script. Additionally, while voice selection is beyond the scope of SpeakEasy, users do have the option of selecting between a male or female voice in the tool (Figure \ref{fig:interface}C). We discuss considerations for approaching identity in voice selection in Section \ref{sec:identity}.

Simon adds ``sound very excited'' as his context and hits generate. SpeakEasy generates a performance of his script tailored to this context and plays back this performance in the Speech Editing View (Figure \ref{fig:interface}.2). 
The Speech Editing View breaks up Simon's script into sentences and displays them in the Sentence Panel. Simon can play the generated speech for the whole script using the playback bar (Figure \ref{fig:interface}G) or play an individual sentence using the show-on-hover Play button next to that sentence (Figure \ref{fig:interface}H, left). SpeechEasy enforces a 0.4 second pause between each sentence. As the speech plays, the spoken word is highlighted.

The Speech Editing View also shows the context (Figure \ref{fig:interface}E) for the current generation. If Simon decides to change his context, he can return to the Script Context View using the Edit Context button (Figure \ref{fig:interface}D). He can also cycle through his changes (D3) using the Undo and Redo buttons (Figure \ref{fig:interface}F).

\subsection{Sentence Level Iteration}

Simon is satisfied with this generation overall, but wants to change how the fourth sentence (Figure \ref{fig:interface}H) is spoken. He clicks on the show-on-hover Edit button (Figure \ref{fig:interface}H, right) for that sentence to reveal the Sentence Iteration menu (Figure \ref{fig:interface}.3). The Sentence Iteration menu provides multiple editing features that ensure a wide variety of performances (D3):

\begin{itemize}
\item \textbf{Speed.} Using the speed slider (Figure \ref{fig:interface}I), users can speed up or slow down the sentence (D5). SpeakEasy offers the following speed options: .75x, .875x, 1x, 1.125x and 1.25x.

\item \textbf{Adjective Recommendations.} Users can also regenerate the sentence performance in the style of an adjective (D2) from the recommended adjectives list (Figure \ref{fig:interface}J).
SpeakEasy provides three distinct recommendations. The first two adjectives are chosen to be relevant and suitable for the sentence being revised. The last adjective is chosen specifically to juxtapose the first two; this last adjective introduces an unanticipated idea, encouraging users to explore divergent thinking and expand their creative horizons (D4). SpeakEasy recommends adjectives that differ from the original generation, thus allowing users to discover different variations that align with their preferences (D3).

\item \textbf{Freeform Text Input.} If the user has a specific change in mind that is not captured by the recommended adjectives, they can type their their own adjective or phrase (D2) into the ``Something else'' text-box (Figure \ref{fig:interface}K) to regenerate. 

\item\textbf{Surprise Take.}
Users can also randomly explore the space of possible generations (D4) by clicking the ``Surprise me with a new take!'' button (Figure \ref{fig:interface}L). SpeakEasy will generate a new performance that is different from any of the previously generated takes for this sentence (D3).
\end{itemize}

Simon iterates on this sentence twice by first selecting ``incredulous'' from the recommended adjectives list and then typing in ``hungry'' into ``Something Else'' text-box.

\subsection{Comparing takes}

 With each iteration, a new tab representing that iteration appears in the takes list below the sentence (Figure \ref{fig:interface}M). The ``hungry'' tab representing the active take is highlighted in green, while tabs for the other iterations (``incredulous'' and ``original'') are greyed out. Simon clicks on the tabs to easily switch between and compare different iterations (D3).

\vspace{6pt}
\noindent 
Once Simon is happy with the entire script, he downloads (Figure \ref{fig:interface}N) the generated performance.

%% file: sections/6_backend_implementation.tex
\section{Implementation}
We develop SpeakEasy as a Wizard-of-Oz system; rather than dynamically generating speech for each user iteration, we simulate the TTS performances by recording multiple performances for a single script. Authors recorded and assigned descriptions to a total of 23 performances. We then processed these recordings with Eleven Lab's speech-to-speech tool\cite{SpeechToSpeech} and a time-stretching algorithm\cite{malah1979time} to yield 230 converted takes of these performances in two anonymized voices (one male, one female) and five speeds. These augmentations enable our voice selection and speed change features. Additionally, we leverage the performance descriptions alongside OpenAI GPT 4 Turbo to support the contextual input and high-level sentence editing features. We chose a Wizard-of-Oz method for two reasons: 1) to guarantee a wide variety of performances more effectively and efficiently and 2) to study high-level context as a potential TTS input mechanism that isn't currently supported by existing TTS algorithms.

\subsection{Creating Performances}

\subsubsection{Writing the Script}
\label{writing}
We drafted an advertisement script compelling listeners to patronize a fictional pizza restaurant called Sandy's (Appendix~\ref{appendix-Script}).
We designed the script to require an expressive performance. The script contains three sections: the hook, the body, and the call to action \footnote{https://blog.vmgstudios.com/how-to-write-video-script}. Each section needs a different delivery to be effective: the hook grabs the listener's attention; the body informs the listener about the product (i.e., a dollar slice pizza); and the call to action encourages them to buy the product.
%As we iterated on the script, we made test recordings and processed them with the speech-to-speech conversion tool. We discovered that the conversion tool yielded artifacts for certain phrases and updated the script accordingly.

\subsubsection{Recording Performances} 
Once we finalized the script, two authors, including one with improv acting experience, collaborated on a joint session to record approximately 30 versions of the script with professional equipment. To ensure a wide variation of performances, the authors improvised both plausible and inventive, off-the-wall interpretations. One author then listened to and labeled each recording with an ID, defining adjective, and a description. These labels are available at Appendix~\ref{appendix-labels}.

\textbf{Failed approaches.} We originally tried to record in a more structured manner by using the affective model of emotion to parameterize our performances \cite{russell1980circumplex}. However, this process yielded performances that were not very nuanced nor diverse and stifled creativity towards inventive performances. We also tried to generate performances using existing TTS tools in lieu of recording them ourselves. However, we were unable to generate a diverse set of varied performances without significant effort.

\subsubsection{Converting Takes}

We upsampled the performances with Adobe's EnhanceSpeech \cite{enhance-speech} to improve audio quality. We then converted the upsampled performances into two voices, one male (Brian) and one female (Jessica) using ElevenLab's speech-to-speech tool. We discarded any performances which contained artifacts from the conversion process, leaving us with 23 performances and 46 converted takes (two takes per performance). Each take was then time-stretched using the audiostretchy python library \footnote{https://pypi.org/project/audiostretchy/}. This time-stretching algorithm preserves pitch to ensure that the sped-up and slowed-down takes still sounded natural. We support 5 speeds for each performance (.75x, .875x, 1x, 1.125x and 2x) yielding 115 converted takes for each voice (230 total). We time-aligned each take to the script using WhisperX \cite{bain2023whisperx} to get sentence and word level timestamps.

We pass along the following information into SpeakEasy:

\begin{itemize}
    \item[1.] \textbf{Performance Data.} A list of 23 performance objects. Each object includes an ID, a defining adjective, and a description.
    \item[2.] \textbf{Converted Takes Data.} A list of 230 converted take objects. Each object includes the audio file, time-aligned transcript, take speed, speaking voice and the performance ID.
\end{itemize}

\subsection{Selecting Takes from Freeform User Input}
\label{sec:take-selection}
Given the user's specification (i.e., the input context for script-level take selection or the sentence descriptor for sentence-level take selection) and the Performance Data, we apply zero shot prompting to ask GPT-4 Turbo (1106-Preview model) to pick the performance ID and description that best matches the user's requirements. Given the selected performance ID, we find the associated converted take that also meets the user's voice and speed specifications. The exact prompts can be found Appendix ~\ref{prompt-context}.

\subsection{Recommending Adjectives}

Adjective recommendations occur after script-level performance selection. For a given sentence, we query GPT-4 with the sentence text and the list of defining adjectives from Performance Data and ask GPT-4 to pick three unique adjectives recommendations from this adjectives list. The adjective associated with the script-level performance is not considered. The first two adjectives should be plausible but creative ways an actor may perform that sentence while the third adjective should be a surprising selection that contrasts the first two. The exact prompt can be found in Appendix~\ref{prompt-adjectives}. In the UI, users may select a recommended adjective multiple times for a single sentence. On the first selection, we return the performance directly associated with that adjective. On subsequent selections, we use the sentence-level take selection method described in ~\ref{sec:take-selection} to pick a closest performance matching that adjective, omitting any previously used performances. 

%% file: sections/7_user_evaluation.tex
\section{User Evaluation}
To understand how SpeakEasy helps users to create expressive speech performances with TTS, we conducted a within-subjects study with twelve participants comparing SpeakEasy to two baselines: ElevenLabs and Speechify. Participants were asked to generate expressive speech for the same script using each of the interfaces. After finishing the generation with each system, participants evaluated the system in questionnaires that included both standard NASA-TLX and design goal specific questions. We also conducted semi-structured closing interviews to understand participant experiences across systems.

\subsection{Hypotheses}
In this study, we investigate the following hypotheses: 

\begin{enumerate}[label=\textit{H\arabic*}]
    \item[H1] SpeakEasy generates significantly \textbf{more suitable performances on the first generation (H1)} than the baselines.
    \item[H2] SpeakEasy generates \textbf{significantly wider variety of performances (H2a)} while being significantly \textbf{more helpful for comparing performances (H2b)} than the baselines.
    \item[H3] SpeakEasy is significantly \textbf{more helpful for modifying performances (H3)} than the baselines.
    \item[H4] SpeakEasy exposes users to significantly \textbf{more creative ideas (H4)} than the baselines.
    \item[H5] SpeakEasy significantly \textbf{increases performance (H5a)} over the baselines while \textbf{decreasing mental demand (H5b), temporal demand (H5c), frustration (H5d), and effort (H5e)} significantly more than the baselines.
    \item[H6] SpeakEasy is significantly \textbf{more useful (H6a), easier to use (H6b), and easier to interpret (H6c)} than the baselines. 
    \item[H7] SpeakEasy's overall speech is \textbf{not significantly more natural than ElevenLabs (76a)} but overall generates significantly \textbf{better performances overall (H7b)} than both baselines.
\end{enumerate}

\subsection{Participants and Procedures}

We recruited twelve participants (Appendix~\ref{appendix-eval-participants}) from our organization through mailing lists and discussion channels. All participants work in various roles related to media creation (e.g., designers, product managers) and identified as either professional(2) or hobbyist(8) creators of speech heavy content. We intentionally recruited for variation in order to garner a diversity of perspectives. Specifically, we wanted to test how the design goals that we created for novice creators could apply towards professionals with more technical knowledge.  

The study sessions were conducted remotely. For each of the three tools (SpeakEasy, ElevenLabs and Speechify), participants were given the pizza commercial script (Section \ref{writing}) and asked to use the tool to generate a performance of that script that would attract customers to the restaurant. They were encouraged to iterate using whatever controls or strategies they saw fit with the goal of getting the speech close to their personal standard.
We counter-balanced the order in which participants used the interfaces. Before each task, we used an example script to guide participants through the features of each tool. Per interface, participants were able to select from a reduced list of one male, and one female voice. We limit voice selection so that participants can more directly compare SpeakEasy (which only has two voice options) with the baselines. Participants were then given up to 7 minutes to complete each task and were encouraged to think aloud. During the closing interview, we asked participants about ways in which each interface helped or frustrated them and strategies they developed. The entire study lasted one hour and participants were compensated \$30 USD.

%% file: sections/8_results.tex
\section{Findings}

After using each interface, participants were asked a series of 5-point Likert scale questions about their experience (Table~\ref{table:results}). These questions evaluate our design goals (notably first generation suitability, variety, comparability, steerability and exposure to creative ideas) and measure task load, usability and speech quality for each system. The precise questions asked are available in Appendix ~\ref{appendix-questions}. We chose not to explicitly evaluate playback speed since we did not contribute novel interactions to it. However, we do report our observations around playback speed when discussing steerability. We summarize the key findings in this section.

\begin{figure*}[h]
  \centering
  \includegraphics[width=\linewidth]{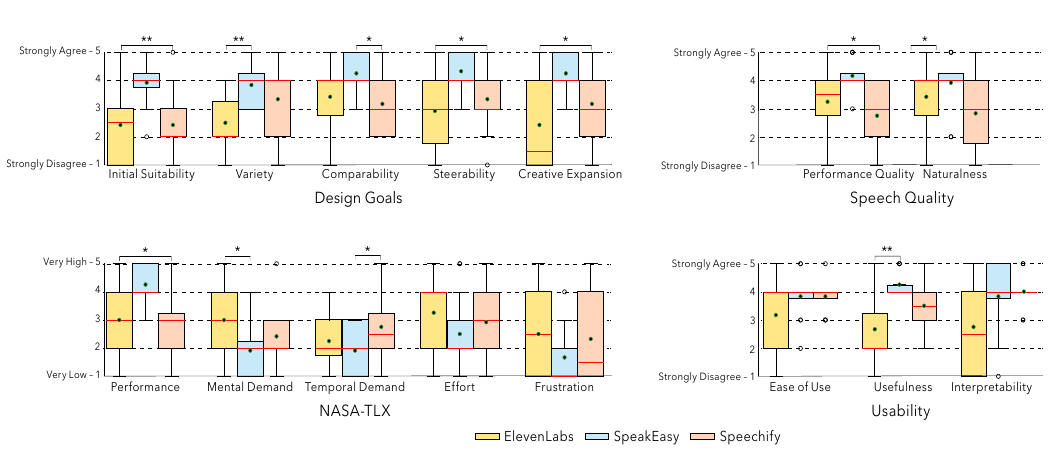}
  \caption{We use a Wilcoxon-Signed Rank Test to evaluate the significance of results comparing SpeakEasy (blue; middle boxes) with Speechify (salmon; right boxes), and ElevenLabs (yellow; left boxes) where significant p-values are indicated by brackets, and stars above the box plots (\( -: p > .100 \), \( +: .050 < p < .100 \), \( *: p < .050 \), \( **: p < .010 \), \( ***: p < .001 \)). Full brackets indicate significant results over both baseline interfaces, half brackets indicate significant results over one baseline interface. Dots are the mean rating, red lines are the median, box heights are interquartile range (IQR),  whiskers correspond to the highest and lowest datum within 1.5 times Q3 and Q1, respectively. Datum outside of the whiskers are labeled as circles. Performance is preferable when higher and has an inverted scale from the rest of the NASA-TLX metrics which are preferable when lower.}
  \label{fig:results}
  \Description{This figure consists of three sets of boxplots comparing user ratings across three systems: ElevenLabs, SpeakEasy, and Speechify. This figure is a visualization of the statistics shared in Table \ref{table:results} which can be cross referenced with this description for full comprehension. The evaluation criteria are categorized into Design Goals, Speech Quality, NASA-TLX Workload Ratings, and Usability. Statistically significant differences are marked with asterisks, indicating where systems diverge in user perceptions. For Design Goals, the metrics include Initial Suitability, Variety, Comparability, Steerability, and Creative Expansion. Speech Quality is evaluated using Performance Quality and Naturalness. NASA-TLX Workload Ratings cover Performance, Mental Demand, Temporal Demand, Effort, and Frustration. For Usability, the metrics include Ease of Use, Usefulness, and Interpretability.}
\end{figure*}

\subsection{Participant Workflows}

\subsubsection{SpeakEasy.}
When generating an initial performance, all but one participant specified an Additional Context; we discuss the participants reactions to the context in Section \ref{section:additional_context}. 
After listening to the initial delivery, one participant regenerated the entire performance by modifying the Additional Context. The remaining participants focused on changing the delivery of specific sentences. Most of these participants explored first using the adjectives recommendations and manually filled out the ``something else'' text box if the recommendations did not meet their needs. Some participants leveraged the ``Surprise Me'' button when they did not want to think of a new textual description. All participants actively used the comparison tabs to pick between takes.

\subsubsection{ElevenLabs.}
All participants generated a initial performance using the default slider values. Participants who were unhappy with this initial result tweaked the sliders in an exploratory manner until they achieved a satisfactory performance. In particular, they changed the stability slider when trying to get more varied outputs and the style exaggeration slider to get more outlandish performances. Some participants added punctuation to the script in order to try to enforce pacing. Many participants generated multiple takes and then referred to the history tab to compare and select a favorite take.

\subsubsection{Speechify.}
To generate an initial performance, half of the participants proactively split the script into one sentence per block while the remainder simply pasted the entire script into one block. 
These latter participants did eventually also split the script to get more control while iterating.
To edit the speech, participants primarily leveraged the pitch slider (to make the voice deeper or higher), speed and pause controls (to changing the pacing) and tone controls (to try to achieve an emotional delivery), though many participants struggled to hear a noticeable difference from changing the tone sliders. Participants with professional editing experience used the timeline to get precise control over pacing and word timings.

\subsection{Effectiveness of Additional Context}
\label{section:additional_context}

\begin{table*}[]
\centering
\begin{tabular}{llccccccccccc}
\toprule
\textbf{Category} & \textbf{Factor} & \multicolumn{2}{c}{\textbf{SpeakEasy}} & \multicolumn{4}{c}{\textbf{ElevenLabs}} & \multicolumn{4}{c}{\textbf{Speechify}}\\
\cmidrule(r){4-5} \cmidrule(r){6-9} \cmidrule(r){10-13} 
 & & \textbf{$\bar{x}$} & \textbf{SD} & \textbf{$\bar{x}$} & \textbf{SD} & \textbf{p} & \textbf{Sig.} & \textbf{$\bar{x}$} & \textbf{SD} & \textbf{p} & \textbf{Sig.} & \textbf{Hypotheses} \\
\midrule
\multirow{5}{*}{\textbf{Design Goals}} 
 & Initial suitability ↑ & \textbf{3.92} & 0.86 & 2.42 & 1.26 & 0.007 & ** & 2.42 & 1.11 & 0.006 & ** & H1 accepted \\
 & Variety ↑ & \textbf{3.83} & 0.8 & 2.5 & 0.96 & 0.007 & ** & 3.33 & 1.25 & 0.329 & - & H2a partially accepted \\
 & Comparability ↑ & \textbf{4.25} & 0.6 & 3.42 & 1.11 & 0.101 & - & 3.17 & 1.21 & 0.019 & * & H2b partially accepted \\
 & Steerability ↑ & \textbf{4.33} & 0.62 & 2.92 & 1.44 & 0.021 & * & 3.33 & 1.11 & 0.028 & * & H3 accepted \\
 & Creative expansion ↑ & \textbf{4.25} & 0.72 & 2.42 & 1.55 & 0.012 & * & 3.17 & 1.21 & 0.02 & * & H4 accepted \\
\midrule
\multirow{5}{*}{\textbf{Task load}} 
 & Performance ↑ & \textbf{4.25} & 0.6 & 3.0 & 1.29 & 0.037 & * & 2.83 & 1.07 & 0.007 & * & H5a accepted \\
 & Mental demand ↓ & \textbf{1.92} & 0.95 & 3.0 & 1.22 & 0.037 & * & 2.42 & 1.04 & 0.058 & + & H5b partially accepted \\
 & Temporal demand ↓ & \textbf{1.92} & 0.86 & 2.25 & 0.92 & 0.305 & - & 2.75 & 1.09 & 0.013 & * & H5c partially accepted \\
 & Effort ↓ & \textbf{2.5} & 1.12 & 3.25 & 1.30 & 0.208 & - & 2.92 & 1.32 & 0.272 & - & H5d rejected \\
 & Frustration ↓ & \textbf{1.67} & 0.94 & 2.5 & 1.44 & 0.203 & - & 2.33 & 1.60 & 0.233 & - & H5e rejected \\
\midrule
\multirow{3}{*}{\textbf{Usability}} 
 & Usefulness ↑ & \textbf{4.25} & 0.43 & 2.67 & 1.11 & 0.009 & ** & 3.5 & 0.96 & 0.064 & + & H6a partially accepted \\
 & Ease of Use ↑ & \textbf{3.83} & 0.8 & 3.17 & 1.4 & 0.203 & - & \textbf{3.83} & 0.55 & 0.931 & - & H6b rejected \\
 & Interpretability ↑ & 3.83 & 1.21 & 2.75 & 1.53 & 0.095 & + & \textbf{4.0} & 0.58 & 0.671 & - & H6c rejected \\
\midrule
\multirow{2}{*}{\textbf{Speech Quality}} 
 & Naturalness ↑ & \textbf{3.92} & 0.95 & 3.42 & 1.32 & 0.301 & - & 2.83 & 1.34 & 0.046 & * & H7a accepted \\
 & Performance quality ↑ & \textbf{4.17} & 0.55 & 3.25 & 1.09 & 0.046 & * & 2.75 & 1.09 & 0.007 & ** & H7b accepted \\
\bottomrule
\end{tabular}
\caption{We use a Wilcoxon-Signed Rank Test to evaluate the significance of results comparing SpeakEasy with Speechify, and ElevenLabs where the p-values (\( -: p > .100 \), \( +: .050 < p < .100 \), \( *: p < .050 \), \( **: p < .010 \), \( ***: p < .001 \)) are reported. The criteria for acceptance is outperforming both baselines with at least \( *: p < .050 \). We also label partial acceptance where we outperform one baseline with \( *: p < .050 \). All other hypotheses not fitting either of these criteria are rejected. Arrows in factor column indicate whether a high or low rating is preferable and the best mean rating is bolded.}
\label{table:results}
\Description{The first column gives the broad category of each of the subjective ratings being tested. The second column lists all subjective ratings. The third column indicates whether a high or low rating is preferable with an arrow icon. Below SpeakEasy, there are two columns pertaining to SpeakEasy's mean and standard deviation for each of the subjective ratings. Below ElevenLabs, there are four columns for its mean, standard deviation, p value with respect to SpeakEasy's results, and a symbol signifying the degree of significance. This four column format is exactly repeated below Speechify. Finally, we include a hypotheses column detailing if our hypotheses have been accepted.}
\label{tab:performance}
\end{table*}

Participants rated SpeakEasy's first generations significantly ($p=.007$ for ElevenLabs and $p=.006$ for Speechify) more suitable than both ElevenLabs and Speechify (H1 accepted). P3, who asked SpeakEasy to perform the script ``like a radio announcer'' noted that SpeakEasy ``understood the prompt perfectly''  
and appreciated ``there was a prompt element to SpeakEasy because a lot of the pre-canned sentiments in the other tools might not be what I'm looking for''. In comparison, most participants felt that the first Speechify and ElevenLabs generations were unusable ($\bar{x}$=2.42, SD=1.11 for Speechify, $\bar{x}$=2.42, SD=1.26 for ElevenLabs). However, one participant (P8) rated ElevenLabs a 5 because ``ElevenLabs just got it pretty close right off the bat.''

Nine participants were able to write an initial context without issue, but the remaining three (P1, P4, P5) struggled with the open-ended prompt. P1 noted that ``explaining with words what you want as a prompt, it's hard'', and appreciated that SpeakEasy suggests a context if the user cannot provide one.

\subsection{Variety and Comparability of Performances}
The variety of SpeakEasy's performances is rated as significantly higher than ElevenLabs ($p=0.007$), but not Speechify ($p=0.329$), (H2a partially accepted). Participants had difficulty generating noticeable variety using ElevenLabs' slider controls. In contrast, most participants commented that both SpeakEasy and Speechify yielded a wide variety of performances, but some participants felt that they could create more ``insane sounding'' takes using Speechify's myriad of controls (e.g., P11 increasing pitch and speed to create an Alvin and the Chipmunks effect). 

Despite having rated Speechify's variety higher, P1 shared in the interview that SpeakEasy's variety felt broadest because ``whenever I chose one of the tags that I had, there was really obvious, like, it was a different performance.'' Consequently, SpeakEasy was rated as significantly more helpful for comparing different takes over Speechify ($p=0.019$), but not ElevenLabs ($p=0.101$), (H2b partially accepted). Both SpeakEasy and ElevenLabs provide features for users to revist past iterations: SpeakEasy using the per sentence tab list and ElevenLabs using a global history. This comparison enabled participants to evaluate whether a change resulted in a better performance and they appreciated being able to reference or revert to a previous iteration if desired. In the interviews, multiple participants shared that they favored SpeakEasy's comparison interface over ElevenLabs because they ``hate going to a different tab'' (P1, P12) to access ElevenLab's history and comparing sentences was easier than comparing whole scripts (P4, P5, P6).

\subsection{Steerability}
\label{section:steerability}
Participants rate SpeakEasy as significantly more helpful than both Speechify ($p=0.028$) and ElevenLabs ($p=0.021$) for modifying performances ($\bar{x}$=3.92, SD=0.86), (H3 accepted). P10 appreciated the flexibility of the ``Something else'' text-box (Figure ~\ref{fig:interface}K), commenting that ``I like being able just to type in what I am imagining.'' P2 and P11 liked both the adjective recommendation (Figure ~\ref{fig:interface}J) and surprise take (Figure~\ref{fig:interface}L) features because these features ``reduce that mental tedium'' and give participants a ``creative mental break.''  In contrast, when using ElevenLabs and Speechify, participants had to both figure out (1) what their goals were at each iteration and (2) how to translate those high level goals into slider values or quantitative features. 

Several participants also liked being able to edit sentences line-by-line in both SpeakEasy and Speechify.  P2 expressed that it's ``so much easier'' when SpeakEasy automatically splits the script compared to the manual blocking in Speechify. Additionally, while we did not explicitly test the playback speed feature, we observed that almost all participants used this functionality across both SpeakEasy and Speechify. Some participants like P1 even requested more granular timing controls in SpeakEasy.

\subsection{Expanding Creative Horizons}
\label{section:expanding}

Participants agreed that SpeakEasy exposed them to creative ideas ($\bar{x}$=4.25, SD=0.72) to a significantly higher degree than ElevenLabs ($p=0.012$) and Speechify (p=$0.02$), (H4 accepted). Six participants shared that SpeakEasy's surprise take and contrasting adjective recommendation features helped them find unexpected, but fitting performances. P4 describes using SpeakEasy as ``it's almost like you have different actors come in and do different auditions and you're like, oh, I wouldn't have even thought of that way of speaking.''

\subsection{Task Load}
Overall, our participants rate that they are more significantly more successful at generating an expressive performance that matches their personal standard when using SpeakEasy ($\bar{x}$=4.25, SD=.6) compared to ElevenLabs ($p=0.037$) and SpeakEasy ($p=0.007$) (H5a accepted). Furthermore, SpeakEasy requires significantly less mental demand ($\bar{x}$=1.92, SD=0.95) than ElevenLabs ($p=0.047$, H5b partially accepted) and and significantly less temporal demand ($\bar{x}$=1.92, SD=0.86) than Speechify ($0.013$, H5c partially accepted). Participants expressed that SpeakEasy's diverse recommendations helped to reduce their mental demand whereas the trial and error attempts to impose local control or manipulate tone with ElevenLabs greatly increased their mental demand. Additionally, compared to SpeakEasy, where users could quickly iterate with high level feedback, or ElevenLabs, where users tweaked a few global control sliders, each iteration in Speechify required adjusting multiple control settings, increasing temporal demand. 

Ratings for other task-load measure such as effort ($\bar{x}$=2.5, SD=1.12 SpeakEasy; $\bar{x}$=3.25, SD=1.3 ElevenLabs; $\bar{x}$=2.92, SD=1.32 Speechify) and frustration ($\bar{x}$=1.67, SD=0.94 SpeakEasy; $\bar{x}$=2.5, SD=1.44 ElevenLabs; $\bar{x}$=2.33, SD=1.6 Speechify) are both lower for SpeakEasy than Eleven Labs and Speechify, but not to a degree of statistical significance (H5d and H5e rejected).
We speculate that this effort and frustration may be due to challenges in open-ended prompting and propose approaches for improving this experience in Section~\ref{sec:prompt-exploration}.

\subsection{Usability of the System}
Users rated SpeakEasy as significantly more useful ($\bar{x}$=4,25, SD=0.43) for generating expressive performances than ElevenLabs ($p=0.009$), and more useful than Speechify, but with marginal significance ($p=0.64$). Thus H6a is partially accepted on behalf of ElevenLabs.
However, SpeakEasy was not significantly easier to use than either baseline and not more interpretable than Speechify. Speakeasy was rated as more interpretable than ElevenLabs with marginal significance. Consequently, H6b and H6c are rejected.
Notably, we had two professionals in our study as well as a few hobbyists who have experience using audio editing tools such as Audition\cite{AdobeAudition}. These participants felt very comfortable with Speechify’s technical controls and even expressed desire for some of these technical controls (e.g., pitch, timeline, pause) in SpeakEasy and ElevenLabs.
In Section~\ref{sec:accomodating-exports}, we discuss how we can accommodate these more expert users while still supporting novice workflows.

\subsection{Speech Quality}
\label{sec:speech-quality}
Participants rate SpeakEasy's performances as significantly higher in quality ($\bar{x}$=4.17, SD=0.55) than both Speechify ($p=0.046$) and ElevenLabs ($p=0.007$) (H7b accepted). Additionally, as we hypothesized in H7a, the naturalness of SpeakEasy performances was rated as better than Speechify ($p=0.46$), but not ElevenLabs. 
We speculate that three factors contribute to SpeakEasy's lack of significant naturalness compared to Eleven Labs. First, while SpeakEasy's performances were manually recorded as part of the Wizard of Oz experiment, the speech-to-speech processing we applied to change the voice did yield slight artifacts. Secondly, SpeakEasy's sentence-by-sentence playback functionality was implemented using HTML5 audio in a non-precise way. This playback sometimes yielded a premature cutoff at the end of a sentence, or a late cutoff that included a small part of the next sentence. Some participants were able to hear these splicing artifacts in SpeakEasy. Finally, ElevenLabs' speech quality is truly impressive. Five participants voted ElevenLabs as generating the most natural speech with multiple participants expressing wonder at ``how natural it sounded.''

%% file: sections/9_discussion_and_future_work.tex
\section{Discussion and Future Work}

\subsection{Limitations and Confounding Factors}

In our investigation, we adopted the Wizard-of-Oz methodology to simulate the TTS capabilities that were not yet developed or widely available at the time of experimentation.
While this approach enabled us to identify an improved user experience, it also introduced certain limitations.

\subsubsection{Use of Human Speech}
The use of human speech in SpeakEasy could confound whether its higher performance rating stems from the system design or the potential quality gap between natural and generated voices. We attempted to mitigate this effect by using speech-to-speech conversion of our natural voices to a synthetic voice provided by ElevenLabs. This conversion reduced the naturalness of our recordings and introduced generated speech artifacts as noted in Section~\ref{sec:speech-quality}. We further sought to evaluate the impact of this factor by asking participants to assess the interfaces for perceived naturalness of speech. 
Our findings indicated that the perceived naturalness from SpeakEasy was not significantly higher than that of ElevenLabs, suggesting that SpeakEasy's performance rating may be attributed to its increase in controllability and the variation embedded in the vocal performances, rather than voice quality alone. Interestingly, there were instances where participants were immediately satisfied with ElevenLab's outputs, indicating that users can achieve high satisfaction with synthetic speech. Speechify received the lowest speech quality rating, likely due to its reliance on traditional audio processing techniques to control the speech generation.

\subsubsection{Applicability of Design Recommendations}
It's worth noting that TTS technology is rapidly evolving, with recent breakthroughs such as GPT-4o Voice~\cite{GPT4o} and other prompt-based TTS models\cite{guo2022promptttscontrollabletexttospeechtext, leng2023prompttts2describinggenerating, liu2023promptstylecontrollablestyletransfer} demonstrating the potential to soon realize systems like SpeakEasy. Such advancements could mitigate many of the limitations presented and lead to the timely application of SpeakEasy's design recommendations. Moreover, even without technological advancements, several of our design recommendations are immediately applicable towards improving the user experience in existing TTS interfaces like ElevenLabs and Speechify. For example, ElevenLabs can integrate speed control and sentence-level iteration while Speechify can implement automatic text blocking and recommendations of quantitative control combinations to reduce user load. Both interfaces could implement mechanisms for easier comparison of performances.

\subsection{Enhancing TTS Interactions for Expressive Content Creation}

The results of our final study show that our participants are significantly more successful at creating an expressive performance matching their personal standard with SpeakEasy than they are with ElevenLabs or Speechify. We attribute this to SpeakEasy effectively reifying our design goals by: giving participants performances tailored to an initial context (D1; H1 accepted); offering high level control for users to modify their performances at a sentence level (D3; H3 accepted); and giving users unanticipated performances that expand their creative horizons (D4; H4 accepted). SpeakEasy is also shown to be give better performances which are not significantly more natural than ElevenLabs (H7 accepted). This provides further confirmation that SpeakEasy's efficacy can be attributed to the features and design goals mentioned above, and the quality of acting in the performances we recorded, not only because of the improved naturalness that comes with human speech. Additionally, SpeakEasy is also rated as significantly more useful than ElevenLabs (H6a partially accepted), while requiring less mental demand than ElevenLabs (H5b partially accepted), and less temporal demand than Speechify (H5c partially accepted). However, there were other metrics for which SpeakEasy did not perform as hypothesized (e.g., variety, effort, ease of use). We discuss potential improvements to SpeakEasy to address these gaps.

\subsubsection{Increasing Variety}
SpeakEasy does not significantly outperform Speechify with respect to variety (H2a rejected w.r.t. Speechify). Our wizard-of-oz design partially contributes to this result because we were limited by the 23 takes that we pre-recorded. However, Speechify does a great job of offering many possible combinations of discrete quantitative controls (e.g., pitch, tone, pause length) to increase variety. For a future iteration of SpeakEasy which is not a Wizard-of-Oz, we can imagine borrowing from Speechify's approach by automatically applying various combinations of quantitative controls to a generation to increase the system's variety instead of relying on the TTS model alone. Such an approach would require additional investigation into how different quantitative changes can map onto high level human intentions. 

Additionally, it could be helpful to train underlying TTS models to explicitly afford absurd, or non-human voices such as the ``chipmunk voice'' created by P11. Users may want to create robotic voices or voices with exaggerated qualities for sci-fi or fantasy contexts, or just for entertainment purposes. 

\subsubsection{Accommodating Expert Editing}
\label{sec:accomodating-exports}

Some study participants, including two professionals, have previously used expert audio editing tools in their creation process. As a result, they felt that Speechify was fairly interpretable and easy to use. These participants recommended that we incorporate Speechify's detailed audio editing features into SpeakEasy. While they appreciated the ability to quickly explore and iterate on generations with SpeakEasy, there were instances when they wanted to make a specific change or last mile improvement (e.g., trimming a pause) and felt it would be easier to make that change with manual editing over by high-level prompting.

In order to support these manual adjustments, while also retaining the beneficial, high-level iterative processes, we propose to add quantative audio controls as secondary features to SpeakEasy. For example, we can keep the current controls on the Sentence Iteration Menu, but add an ``advanced controls'' button that would expand the menu to include the quantitative audio controls. In this way, we can integrate these expert features for users who seek them out without overwhelming users who don't.

\subsubsection{Enabling Prompt Exploration}
\label{sec:prompt-exploration}

Some participants struggled to write a context for the script performance in SpeakEasy. These participants either did not have an idea in mind or had difficulty articulating their creative goals. We propose utilizing methods for interactive prompt exploration, similar to those in Promptify \cite{brade2023promptify}, towards context writing to bootstrap users and help them ideate. 

\subsection{Additional Considerations for TTS Design}

\subsubsection{Affirming Identity with TTS}
\label{sec:identity}
Almost all of the participants from our first comparative evaluation study (Section ~\ref{sec:formative-compare}) tried to affirm their identities by searching for a voice that resembles their own. One participant mentioned that their voice is deeply tied to their racial identity, and that they did not feel represented by the available options. While it's easy to suggest that TTS developers train to accommodate for a wide range of accents, researchers also need to consider ethical questions around how to present and help users search for different voices without stereotyping. Relatedly, two participants (C1, C7) expressed interest in voice cloning as a means to preserve their own identity rather than find a proxy for it. Voice cloning also requires strong ethical considerations around ownership, consent, and privacy. 

\subsubsection{Adapting to additional use cases}
In this work, we focus on short, single speaker speech commonly found in social videos by independent content creators. However, expressive TTS also holds value in many other application settings such as multi-speaker content or long-form (greater than 10 minutes) content (e.g., video essays, audio books). We believe the design goals which we reify in SpeakEasy are still broadly applicable when combined with additional considerations for these scenarios. For example, to adapt to multi-speaker content, we recommend allowing users to specify a context for each speaker, rather than just one context for the whole script. Additionally since multi-speaker dialogue is often reactive, adjective recommendations for a sentence can be further conditioned on the delivery of the previous sentence. For long-form content, it's possible to help users work on smaller sections of the script at a time by using a large language models to split the script based on narrative arc or individual scenes. The TTS generation in each of these sections could then be supported by our design goals. 

\subsection{Recommendations for Machine Learning Researchers}
As SpeakEasy is a Wizard-of-Oz prototype, we make two recommendations to the speech AI community towards making expressive TTS a reality. 

\subsubsection{Allow contextual conditioning of generated speech.}
TTS researchers should continue to develop strategies for allowing users to condition generated speech with additional context. While progress has been made with respect to this suggestion \cite{overview_of_affective} as manifested in Speechify's tone controls, we believe this context should go beyond predetermined emotion sliders. Emotions are nuanced and different people may have different interpretations of an emotion. For example, P3 expressed that ``a lot of the pre-canned sentiments in the other tools might not be what I’m looking for''. Instead, in order to align with expert workflows for voice acting, we recommend allowing for flexible contextual inputs such as character descriptions, character references, exemplar performances, or even more exotic conditioning like video samples.

\subsubsection{Enhance the variety of generated speech.}
Our study participants struggled to generate diverse performances with ElevenLabs and succeeded in Speechify by modifying the generated speech with audio editing features. However, this modification process expended user effort and sometimes resulted in less natural sounding output. TTS researchers should develop models to generate a wide variety of performances for a single script. As described in the voice actor interviews and further supported in the user study findings, providing diverse variations can both compensate for vague user intent and help expand users' creative horizons by giving unexpected but pleasant results. Future work might be able to improve variety by conditioning a model on past iterations and ensuring that new generations diverge significantly from previous ones. We also speculate that the variety gap between human and synthetic speech is a data issue, and that the breadth of performances used to train TTS is too narrow. Future work can focus on trying new strategies for data collection. When actively recording speech for new datasets, we recommend attempting to have subjects improvise multiple interpretations for a script, and labelling the performances after the fact. Such a strategy could increase the novelty of the recordings produced, and potentially widen the variety of speech in a model trained with this data. 

%% file: sections/10_conclusion.tex
\section{Conclusion}
We employed a two pronged approach towards designing TTS interactions for expressive content creation. First, we conduct a comparative evaluation of existing TTS tools with eight independent content creators who record expressive speech for their own videos. This study showed us that some interfaces offer controls that are too granular for easy and effective expressive control while others offer controls which are too difficult to interpret, leading to difficulty iterating.  Next, we conducted expert interviews with eight professional voice actors to understand effective strategies of communicating intention for expressive speech performances. Voice actors typically intake some context (e.g., character description, reference example) that helps them understand how to interpret the script and give an initial performance. Then, they respond to vague feedback from stakeholders by providing a variety of performances. We synthesized these findings from both studies into a set of design guidelines. We validate these guidelines by reifying them into SpeechEasy, a Wizard-of-Oz system for crafting expressive TTS performances. Our final user study validated many of our design goals by showing that participants are more successful at generating expressive performances matching their personal standard when working with SpeakEasy, while requiring no more effort than two industry leading interfaces. We close this work with future directions, highlighting how SpeakEasy's features may be implemented in reality, and how our design goals may be applied to related applications of TTS.

\begin{acks}

We would like to thank Joy Kim, Shm Almeda, Bryan Wang, Ding Li and our reviewers for their constructive feedback that helped us strengthen our exposition. We would also like to thank all the content creators, voice actors and user evaluation participants for their invaluable insights.

\end{acks}

%% file: sections/11_appendix.tex
\clearpage
\onecolumn

\section{Appendix}
\label{appendix}

\subsection{Formative Participants}
\label{appendix-formative}
\begin{table*}[!h]
\begin{tabular}{| c | c | c | c |}
\hline
 & \textbf{Content Type}& \textbf{Script Topic} & \textbf{TTS experience}\\
 \hline
 \textbf{C1}&  Promotional videos for podcasts & Podcast promo & Tried once\\
 \hline
 \textbf{C2}&  Promotional videos for small business & Business tips & Tried once\\
 \hline
 \textbf{C3}&  Video game fan fiction & Video game fan fic & None\\
 \hline
 \textbf{C4}&  Documentaries, explainer videos & Explainer video narration & Tried once\\
 \hline
 \textbf{C5}&  Product promotion TikToks & Product ad & None \\
\hline
\textbf{C6}&  News packages & Explainer video narration & Tried once \\
\hline
\textbf{C7}&  Travel vlogs & Vlog intro & Tried once \\
\hline
\textbf{C8}&  Instructional music videos &  Music instructions & Tried once \\
\hline
\end{tabular}
\caption{Comparative Evaluation Participants. Background information of the content creators. All creators had at least one year of experience making videos for social media. Six of the creators tried TTS once, but none had extensive TTS experience prior to the study.}
\label{table:content_creators}
\Description{This table contains four columns: the first column contains identifiers for content creator participants, the second contains the types of content that they create, the third describes the type of script they used in the formative study, and the fourth describes their level of experience with TTS.}
\end{table*}

\begin{table*}[!h]
\begin{tabular}{| c | c |}
\hline
 & \textbf{Voice Acting Experience} \\
 \hline
 \textbf{V1} & Commercials, internal training videos \\
 \hline
 \textbf{V2} & Gig work, commercials, English to Japanese dubbing \\
 \hline
 \textbf{V3} & Animation, internal training videos \\
 \hline
 \textbf{V4} & Audio-books, animation, commercials \\
 \hline
 \textbf{V5} & Audio-books, e-learning, explainer videos, animation, video games \\
\hline
\textbf{V6} & Narrative podcasts, news packages, commercials \\
\hline
\textbf{V7} & Radio, television, film, commercials \\
\hline
\textbf{V8} & Commercials, video games, narrations, animation \\
\hline
\end{tabular}
\caption{Background information of the voice actors. All voice actors had at least two years of experience doing voice acting for a variety of projects from animation to video games.}
\Description{The first column contains identifiers for the different voice actors in the formative study, and the second describes the sort of voice acting projects they have worked on.}
\label{table:voice_actors}
\end{table*}

\subsection{ElevenLabs Control Parameter Descriptions}
\label{appendix-elevenlabs}
Here we include descriptions of ElevenLabs control parameters taken directly from their website \footnote{\url{https://elevenlabs.io/app/speech-synthesis/text-to-speech}}.

\begin{itemize}
    \item \textbf{Stability:} Increasing stability will make the voice more consistent between re-generations, but it can also make it sound a bit monotone. On longer text fragments, we recommend lowering this value.
    
    \item \textbf{Similarity:} High enhancement boosts overall voice clarity and target speaker similarity. Very high values can cause artifacts, so adjusting this setting to find the optimal value is encouraged.
    
    \item \textbf{Style Exaggeration:} High values are recommended if the style of the speech should be exaggerated compared to the uploaded audio. Higher values can lead to more instability in the generated speech. Setting this to 0.0 will greatly increase generation speed and is the default setting.
\end{itemize}

\subsection{Script}
\label{appendix-Script}

\begin{itemize}
    \item[] \textbf{Hook.} Wait, what? Sandy’s Slices has a dollar slice now? I am on my way.
    \item[] \textbf{Body.} This is the Queen Margherita, with creamy mozzarella, a tangy vodka sauce, and fresh basil. But only one dollar?! Chef’s kiss.
    \item[] \textbf{Call to Action.} Why settle for less when you can get all this for a buck. Get out to Sandy’s now!
\end{itemize}

\subsection{Performance Data}
\label{appendix-labels}

\begin{verbatim}
[{ description: "Abrasive", defining_adjective: "Abrasive", performance_id: 0 },
{ description: "Commanding, Intense", defining_adjective: "Commanding", performance_id: 1 },
{ description: "Enthusiastic, Zesty", defining_adjective: "Enthusiastic", performance_id: 2 },
{ description: "Giggly, Bubbly", defining_adjective: "Giggly", performance_id: 3 },
{ description: "Nasally, Annoyed", defining_adjective: "Nasally", performance_id: 4 },
{ description: "Pensive, taking their time", defining_adjective: "Pensive", performance_id: 5 },
{ description: "Percussive, Flair", defining_adjective: "Percussive", performance_id: 6 },
{ description: "Sing song, cheerful", defining_adjective: "Sing song", performance_id: 7 },
{ description: "Mellow, Friendly", defining_adjective: "Mellow", performance_id: 8 },
{ description: "Slow", defining_adjective: "Slow", performance_id: 9 },
{ description: "Punchy", defining_adjective: "Punchy", performance_id: 10 },
{ description: "Chill, Casual, Laughing", defining_adjective: "Chill", performance_id: 11 },
{ description: "Somewhat hyped", defining_adjective: "Somewhat hyped", performance_id: 12 },
{ description: "Relaxed, close to the mic, NPR", defining_adjective: "Relaxed", performance_id: 13 },
{ description: "Broadcaster, Exaggerated", defining_adjective: "Broadcaster", performance_id: 14 },
{ description: "Laid back, surfer bro", defining_adjective: "Laid back", performance_id: 15 },
{ description: "Fairly enthusiastic, laid back", defining_adjective: "Fairly enthusiastic", performance_id: 16 },
{ description: "Breathy", defining_adjective: "Breathy", performance_id: 17 },
{ description: "Motivating", defining_adjective: "Motivating", performance_id: 18 },
{ description: "Annunciated, Slow, Clear", defining_adjective: "Annunciated", performance_id: 19 },
{ description: "Incredulous", defining_adjective: "Incredulous", performance_id: 20 },
{ description: "Saucy", defining_adjective: "Saucy", performance_id: 21 },
{ description: "Subdued", defining_adjective: "Subdued", performance_id: 22 }]
\end{verbatim}

\subsection{GPT Prompts}

\subsubsection{Selecting a Performance from Freeform User Input.}
\label{prompt-context}
``You are presented with an array of performances and descriptions and are also provided with a query. The performances correspond to a voice actor's performance of a script, and the description explains their performance. The collection is presented as an array of jsons with the json as 'x'. Each 'x' will have a field: 'performance\_id' which is the integer label for a given take, and 'description' which is the description of that take.
Your task is to return three of the performances with the most relevant description to the query by returning a list that looks as follows [<x.take\_id>, <x.take\_id>, <x.take\_id>] where <x.take\_id> is the take\_id value for the relevant take that you have selected.

The query is: <USER CONTEXT>

The performances and descriptions are: <PERFORMANCES FROM Appendix \ref{appendix-labels}>

Please select three takes and descriptions relevant to "<USER CONTEXT>". Make sure to keep the the result semantically meaningful and always select three takes. If there is no good match, select the closest takes. 
THE OUTPUT SHOULD ONLY BE THE LIST OF TAKE IDS WITHOUT ADDITIONAL TEXT. For example, [<x.take\_id>, <x.take\_id>, <x.take\_id>].''

\subsubsection{Recommending Adjectives}
\label{prompt-adjectives}

``You are presented with a line from a script (a string), the whole script (an array of strings), and an array of adjectives (an array of strings). The script is the line currently being performed, and the list of adjectives describe existing performances that can be selected for the user.
Your task is to return three UNIQUE adjectives THAT MUST BE FROM THE ARRAY OF ADJECTIVES.

For the first two adjectives, choose two which you feel are plausible ways that an actor might perform of the script and BE CREATIVE, but make the third adjective a surprising selection which contrasts the first two.
AGAIN, EACH OF THE ADJECTIVES MUST BE FROM THE LIST OF ADJECTIVES PROVIDED AND EACH MUST BE UNIQUE.
Return them like this like this ["adjectives[i]", "adjectives[j]", "adjectives[k]"] where i, j, k are the indices that you have selected.
The line is: <SENTENCE>

The script: <FULL SCRIPT ~\ref{appendix-Script}>

The array of adjectives: <LIST OF DEFINED ADJECTIVES FROM ~\ref{appendix-labels}>

Please select three adjectives: two that fit "<SENTENCE>" and one which contrasts the first two selected adjectives. Make sure to keep the the result semantically meaningful.
THE OUTPUT SHOULD ONLY BE THE ARRAY OF SELECTED ADJECTIVES WITHOUT ADDITIONAL TEXT. For example, ["adjectives[i]", "adjectives[j]", "adjectives[k]"] where i, j, k are the indices that you have selected.''

\subsection{User Evaluation Supplemental Data}

\subsubsection{Participants}
\label{appendix-eval-participants}

Please see Table~\ref{table:user-eval-participants}.

\begin{table*}[ht]
\begin{tabular}{@{}p{1cm} p{3cm} p{1.8cm} p{6cm} p{1cm}@{}}
\toprule
\textbf{ID} &  \textbf{Title} & \textbf{Experience} & \textbf{Content Type} & 
\textbf{Has Used TTS} \\ \midrule
P1 & Designer & Hobbyist & Short-form videos & Yes \\ \midrule
P2 & Video Journalist & Professional & Broadcast videos, reels, explainer \& promotional content  & Yes \\ \midrule
P3 & Strategic Development Manager & Professional & Long-form \& short-form video & Yes \\ \midrule
P4 & Designer & Hobbyist & Product vision videos, short-form art videos & No \\ \midrule
P5 & PhD Student & Hobbyist & Academic paper videos, travel shorts & Yes \\ \midrule
P6 & Creative Director & Hobbyist & Long-form YouTube content with short-form social cut-downs & No \\ \midrule
P7 & Software Engineer & Hobbyist & TikToks, Insta Reels & No \\ \midrule
P8 & Product Manager & Professional & Film and television & Yes \\ \midrule
P9 & Strategic Development Director & Hobbyist & Youtube Videos, films & Yes \\ \midrule
P10 & UX Designer & Hobbyist & Podcasts, short YouTube tutorials & Yes \\ \midrule
P11 & Strategic Development Manager & Hobbyist & TikTok/Instagram short-form videos, promotional content, audio podcast, sports content & No \\ \midrule
P12 & Senior Category Lead & Hobbyist & Instructional videos & No \\ \bottomrule
\end{tabular}
\caption{ This table describes the relevant experiences of participants in our final user study. Professional - (Full-time creator, primary source of income), Hobbyist - (Part-time creator, some income), Novice - (New creator, no income)}
\Description{The first column lists the IDs of the participants in the final user study, the second gives their job title, the third gives their experience-level with content creation, the fourth describes the type of content they created, and the fifth explains whether or not each participant has used TTS before.}
\label{table:user-eval-participants}
\end{table*}

\subsubsection{Questionnaire}
\label{appendix-questions}

\subsubsection*{NASA-TLX Task Load Questions}
\begin{enumerate}[leftmargin=4em, itemsep=0pt]
    \item \textbf{Performance:} How successful were you in accomplishing what you were asked to do?
    \item \textbf{Mental demand:} How mentally demanding was using this interface?
    \item \textbf{Temporal demand:} How hurried or rushed was the pace of the task?
    \item \textbf{Effort:} How hard did you have to work to accomplish your level of performance?
    \item \textbf{Frustration:} How insecure, discouraged, irritated, stressed, and annoyed were you?
\end{enumerate}

\subsubsection*{Usability Questions}
\begin{enumerate}[leftmargin=4em, itemsep=0pt]
    \item \textbf{Usefulness:} The controls on this interface were useful.
    \item \textbf{Ease of Use:} The controls on this interface were easy to interpret.
    \item \textbf{Interpretability:} The controls on this interface were intuitive and easy to use.
\end{enumerate}

\subsubsection*{Design Goals Questions}
\begin{enumerate}[leftmargin=4em, itemsep=0pt]
    \item \textbf{Initial suitability:} My first generated speech matched what I was looking for.
    \item \textbf{Variety:} This system gave a wide variety of generated speech.
    \item \textbf{Comparability:} This system helped me compare a new generated speech with previous iterations.
    \item \textbf{Steerability:} This system helped me modify the generated speech to fit my goals for the voiceover.
    \item \textbf{Creative expansion:} This system exposed me to new creative ideas.
\end{enumerate}

\subsubsection*{Speech Quality Questions}
\begin{enumerate}[leftmargin=4em, itemsep=0pt]
    \item \textbf{Naturalness:} Ignoring the controls of this system, this generated speech was natural.
    \item \textbf{Performance quality:} Ignoring the controls of this system, this generated speech fit my goals for the script.
\end{enumerate}